\documentclass[aps,prl,superscriptaddress,twocolumn,showpacs]{revtex4}

\usepackage{amsmath,amsfonts,bm,graphicx}

\newcommand{\om}{\omega}

\begin{document}

\title{Shot Noise of a Quantum Shuttle}

\author{Tom\'a\v s Novotn\'y}
\email{tno@mic.dtu.dk}

\affiliation{MIC -- Department of Micro and Nanotechnology,
             Technical University of Denmark,
             DTU -- Building 345 East,
             DK-2800 Kongens Lyngby, Denmark}

\affiliation{Department of Electronic Structures,
     Faculty of Mathematics and Physics, Charles University,
     Ke Karlovu 5, 121 16 Prague, Czech Republic}

\author{Andrea Donarini}
%\email{ad@mic.dtu.dk}

\affiliation{MIC -- Department of Micro and Nanotechnology,
             Technical University of Denmark,
             DTU -- Building 345 East,
             DK-2800 Kongens Lyngby, Denmark}

\author{Christian Flindt}
%\email{cf@mic.dtu.dk}

\affiliation{MIC -- Department of Micro and Nanotechnology,
             Technical University of Denmark,
             DTU -- Building 345 East,
             DK-2800 Kongens Lyngby, Denmark}

\author{Antti--Pekka Jauho}
%\email{antti@mic.dtu.dk}

\affiliation{MIC -- Department of Micro and Nanotechnology,
             Technical University of Denmark,
             DTU -- Building 345 East,
             DK-2800 Kongens Lyngby, Denmark}

\date{\today}

\begin{abstract}
We formulate a theory for shot noise in  quantum
nanoelectromechanical systems.  As a specific example, the theory
is applied to a quantum shuttle, and the zero-frequency noise,
measured by the Fano factor $F$, is computed. $F$ reaches very low
values ($F\simeq 10^{-2}$) in the shuttling regime even in the
quantum limit, confirming that shuttling is universally a low
noise phenomenon. In approaching the semiclassical limit, the Fano
factor shows a giant enhancement ($F\simeq 10^2$) at the shuttling
threshold, consistent with predictions based on phase-space
representations of the density matrix.
\end{abstract}


\maketitle Nanoelectromecanical systems (NEMS) are presently a
topic of intense research activity \cite{cra-sci-00}.  These
devices combine electronic and mechanical degrees of freedom to
display new physical phenomena, and potentially may lead to new
functionalities. An archetypal example of such a new phenomenon is
the charge shuttling transition exhibited by the device proposed
by Gorelik et al.~\cite{gor-prl-98}: here a movable nanoscopic
object begins to transport electrons one-by-one beyond a certain
threshold bias.  Recent work has extended the original ideas to
the quantum regime (the motion of the movable part is also
quantized), and has shown that the shuttling transition occurs
even in this limit, albeit in a smeared-out form
\cite{arm-prb-02,nov-prl-03,smi-prb-04,fed-prl-04}.

An unequivocal experimental observation of the shuttling
transition has not yet been achieved.  The IV-curve measured in
the recent experiments on a C$_{60}$ single-electron transistor
can be interpreted in terms of shuttling \cite{fed-epl-02}, but
also alternative explanations have been promoted
\cite{boe-epl-01,mcc-prb-03,braig-prb-03}.  It is therefore
natural to look for more refined experimental tools than just the
average current through the device.  An obvious candidate is the
current noise spectrum \cite{bla-phr-00,bee-pht-03}. The
measurement of the noise spectrum or even higher moments (full
counting statistics) reveals more information about the transport
through the device than just the mean current. The theoretical
studies of the noise have attracted much attention recently in
NEMS in general
\cite{mit-preprint-03,arm-preprint-04,cht-preprint-04} as well as
for the shuttling setup
\cite{pis-preprint-04,isa-epl-04,rom-preprint-04} in the
(semi)classical limit. More specifically, Pistolesi
\cite{pis-preprint-04} reports a vanishing Fano factor in the
large amplitude limit of a driven shuttle. On the other hand, a
study by Isacsson and Nord \cite{isa-epl-04} of a classical system
exhibiting shuttling instability found, somewhat surprisingly, a
higher Fano factor in the shuttling regime than in the tunneling
regime. This result is attributed to a different confining
potential used in \cite{isa-epl-04} compared to the other studies
in this area.

The aim of this Letter is to develop a quantum mechanical theory
for the shot noise spectrum for quantum NEMS, and apply it to the
model introduced by Gorelik et al.\ \cite{gor-prl-98}. Our method
combines the classical nature of the charge transfer processes in
the high bias limit \cite{ela-pla-02} with an operator version of
a generating function technique \cite{vankampen}.  While the
present work considers only Markovian master equations, we believe
that the method can be generalized to the case where the dynamics
of the mechanical degrees of freedom is non-Markovian. For systems
where the current noise can be expressed in terms of {\it system}
operators (using the quantum optics language), such as the quantum
dot array of Ref. \cite{arm-prb-02}, an alternative evaluation of
noise, based on the quantum regression theorem (QRT) is possible,
and we have verified that the two methods give identical results
in this case \cite{fli-preprint-04}.  We stress that the converse
is not true: QRT is not applicable to the single-dot case.

Our previous quantum calculation \cite{nov-prl-03} of the mean
current relied on a generalized master equation (GME) for the
system density matrices $\rho_{ii}(t)$ ($\rho_{11}$ and
$\rho_{00}$ describe the occupied and empty dot, respectively, and
the off-diagonal components decouple from their dynamics, and can
be neglected).  In order to compute the noise spectrum, we follow
the ideas of Gurvitz and Prager \cite{gur-prb-96}, and introduce
number-resolved density-matrices $\rho_{ii}^{(n)}$, where
$n=0,1,\dots$ is the number of electrons tunneled into the right
lead by time $t$. Obviously, $\rho_{ii}(t)=\sum_n
\rho_{ii}^{(n)}(t)$. The $\rho_{ii}^{(n)}$ obey
\begin{equation}\label{GME}
\begin{split}
    \dot{\rho}_{00}^{(n)}(t) &= \frac{1}{i\hbar} [H_{\rm osc},\rho_{00}^{(n)}(t)]
    + \mathcal{L}_{\rm damp}\,\rho_{00}^{(n)}(t)\\
    &\quad-
    \frac{\Gamma_L}{2}\{e^{-2x/\lambda},\rho_{00}^{(n)}(t)\}
    + \Gamma_R e^{x/\lambda}\rho_{11}^{(n-1)}(t)e^{x/\lambda}
    , \\
    \dot{\rho}_{11}^{(n)}(t) &= \frac{1}{i\hbar}[H_{\rm osc}-eEx,\rho_{11}^{(n)}(t)]
    + \mathcal{L}_{\rm damp}\,\rho_{11}^{(n)}(t)\\
    &\quad-
    \frac{\Gamma_R}{2}\{e^{2x/\lambda},\rho_{11}^{(n)}(t)\}
    + \Gamma_L e^{-x/\lambda}\rho_{00}^{(n)}(t)e^{-x/\lambda}
    ,
\end{split}
\end{equation}
with $\rho_{11}^{(-1)}(t)\equiv 0$. In \eqref{GME}, the
commutators describe coherent evolution of discharged or charged
harmonic oscillator of mass $m$ and frequency $\om$ in electric
field $E$, respectively. The terms involving $\Gamma_{L,R}$
describe the charge transfer processes from/to leads while the
mechanical damping with the damping coefficient $\gamma$ is
determined by the kernel (at $T=0$) \cite{nov-prl-03}
\begin{equation}
\mathcal{L}_{\rm
damp}\,\rho=-\frac{i\gamma}{2\hbar}[x,\{p,\rho\}]- \frac{\gamma
m\omega}{2\hbar}[x,[x,\rho]].
\end{equation}

The mean current and the zero-frequency shot noise are given by
\cite{ela-pla-02}
\begin{align}
    I = e &\frac{d}{dt}\sum_nn P_n(t)\Big|_{t\to\infty}
    = e \sum_n n \dot{P}_n(t)\Big|_{t\to\infty},\label{current}\\
    S(0) = 2e^2&\frac{d}{dt}\bigg[\sum_nn^2P_n(t)
    -\Big(\sum_nnP_n(t)\Big)^2\bigg]\bigg|_{t\to\infty}
    \label{macdonald},
\end{align}
where
$P_n(t)=\mathrm{Tr_{osc}}[\rho_{00}^{(n)}(t)+\rho_{11}^{(n)}(t)]$
are the probabilities of finding $n$ electrons in the right lead
by time $t$. Using  Eq.~\eqref{GME} we find $I=\sum_n n
\dot{P}_n(t)
=\Gamma_R\mathrm{Tr_{osc}}\big(e^{2x/\lambda}\rho_{11}(t)\big)$,
i.e. one recovers the stationary current used previously
\cite{nov-prl-03}.   In a similar fashion, $\sum_n n^2
\dot{P}_n(t)
=\Gamma_R\mathrm{Tr_{osc}}\big\{e^{2x/\lambda}\big[2\sum_n n
\rho_{11}^{(n)}(t) + \rho_{11}(t)\big]\big\}$, whose large-time
asymptotics determines the shot noise according to
(\ref{macdonald}). This can be computed using an operator-valued
generalization of the generating function concept. We introduce
the generating functions $F_{ii}(t;z)=\sum_n
\rho_{ii}^{(n)}(t)z^n$ with the properties
$F_{ii}(t;1)=\rho_{ii}(t),\,\tfrac{\partial}{\partial
z}F_{ii}(t;z)|_{z=1}=\sum_n n \rho_{ii}^{(n)}(t)$. The  equations
of motion for $F_{ii}(t;z)$ are
\begin{equation}\label{GMEgen}
\begin{split}
    \frac{\partial}{\partial t}F_{00}(t;z) &= \frac{1}{i\hbar} [H_{\rm osc},F_{00}(t;z)]
    - \frac{\Gamma_L}{2}\{e^{-2x/\lambda},F_{00}(t;z)\}\\
    &\quad+\mathcal{L}_{\rm damp}\,F_{00}(t;z)
    + z \Gamma_R e^{x/\lambda}F_{11}(t;z)e^{x/\lambda}\\
    &=\mathcal{L}_{00} F_{00}(t;z) + z \mathcal{L}_{01}F_{11}(t;z), \\
    \frac{\partial}{\partial t} F_{11}(t;z) &= \frac{1}{i\hbar}
    [H_{\rm osc}-eEx,F_{11}(t;z)]+ \mathcal{L}_{\rm damp}\,F_{11}(t;z)\\
    &\quad- \frac{\Gamma_R}{2}\{e^{2x/\lambda},F_{11}(t;z)\}\\
    &\quad+ \Gamma_L e^{-x/\lambda}F_{00}(t;z)e^{-x/\lambda}\\
    &= \mathcal{L}_{10} F_{00}(t;z) + \mathcal{L}_{11}F_{11}(t;z),
\end{split}
\end{equation}
where we have introduced the block structure of the Liouvillean
(super)operator
$\mathcal{L}=\left(\begin{smallmatrix}\mathcal{L}_{00}&\mathcal{L}_{01}
\\ \mathcal{L}_{10}&\mathcal{L}_{11}\\\end{smallmatrix}\right)$.
Using the $F$'s the shot noise formula can be rewritten as
\begin{widetext}
\begin{equation}\label{noisegen}
\frac{S(0)}{2e^2\Gamma_R}=\Big\{\mathrm{Tr_{osc}}\Big[e^{2x/\lambda}\Big(2\frac{\partial}
{\partial z}F_{11}(t;z)\Big|_{z=1}+F_{11}(t;1)\Big)\Big]
-2\mathrm{Tr_{osc}}\Big[e^{2x/\lambda}F_{11}(t;1)\Big]
\mathrm{Tr_{osc}}\Big[\frac{\partial}{\partial z} \sum_{i=0}^1
F_{ii}(t;z)\Big|_{z=1}\Big]\Big\}\Big|_{t\to\infty}.
\end{equation}
\end{widetext}

A Laplace transform of \eqref{GMEgen} yields
\begin{equation}
    \begin{pmatrix}
    \tilde{F}_{00}(s;z)\\
    \tilde{F}_{11}(s;z)
    \end{pmatrix} =
    \begin{pmatrix}
    s - \mathcal{L}_{00} & -z\mathcal{L}_{01}\\
    -\mathcal{L}_{10} & s - \mathcal{L}_{11}
    \end{pmatrix}^{-1}
    \begin{pmatrix}
    f_{00}^{\rm init}(z)\\
    f_{11}^{\rm init}(z)
    \end{pmatrix}
\end{equation}
where $f_{ii}^{\rm init}(z) = \sum_n \rho_{ii}^{(n)}(0)z^n$
depends on the initial conditions. Defining the resolvent
$\mathcal{G}(s)=(s-\mathcal{L})^{-1}$ of the full Liouvillean we
arrive at
\begin{equation}
    \begin{pmatrix}
    \tilde{F}_{00}(s;1)\\
    \tilde{F}_{11}(s;1)
    \end{pmatrix} = \mathcal{G}(s)
    \begin{pmatrix}
    \rho_{00}^{\rm init}\\
    \rho_{11}^{\rm init}
    \end{pmatrix}, \label{Ffunction}
\end{equation}
\begin{equation}
\begin{split}
    \frac{\partial}{\partial z}\left.\begin{pmatrix}
    \tilde{F}_{00}(s;z)\\
    \tilde{F}_{11}(s;z)
    \end{pmatrix}\right|_{z=1} &=
    \mathcal{G}(s)
    \begin{pmatrix}
      0 & \mathcal{L}_{01} \\
      0 & 0 \\
    \end{pmatrix}
    \mathcal{G}(s)
    \begin{pmatrix}
    \rho_{00}^{\rm init}\\
    \rho_{11}^{\rm init}
    \end{pmatrix}\\
    &\quad+\mathcal{G}(s)
    \begin{pmatrix}
    f_{00}^{'\rm init}(1)\\
    f_{11}^{'\rm init}(1)
    \end{pmatrix} \label{Ffunctder}.
\end{split}
\end{equation}
In order to extract the large-$t$ behavior  we study the
asymptotics of the above expressions as $s\to 0+$. This is
entirely determined by the resolvent $\mathcal{G}(s)$ in the
small-$s$ limit. Since $\mathcal{L}$ is singular (recall
$\mathcal{L}\rho^{\rm stat}=0$) the resolvent is singular at
$s=0$. To extract the singular behavior we introduce the
projector $\mathcal{P}$ on the null space of the Liouvillean:\
$\mathcal{P}\bullet=\left(\begin{smallmatrix}\rho_{00}^{\rm stat}&
\\ \rho_{11}^{\rm stat}\\\end{smallmatrix}\right)\mathrm{Tr_{sys}}(\bullet)$. We
also need the complement $\mathcal{Q}=1-\mathcal{P}$. Using the
relations $\mathcal{P}\mathcal{L}=\mathcal{L}\mathcal{P}=0$ and
$\mathcal{L}=\mathcal{Q}\mathcal{L}\mathcal{Q}$, the resolvent can
be expressed as $\mathcal{G}(s)=(s\mathcal{P}+s\mathcal{Q}
-\mathcal{Q}\mathcal{L}\mathcal{Q})^{-1}=\frac{1}{s}\mathcal{P}+
\mathcal{Q}\frac{1}{s-\mathcal{L}}\mathcal{Q}\approx
\frac{1}{s}\mathcal{P}-\mathcal{Q}\mathcal{L}^{-1}\mathcal{Q}$, in
leading order for small $s$. The object
$\mathcal{Q}\mathcal{L}^{-1}\mathcal{Q}$ (the pseudoinverse of
$\mathcal{L}$) is regular as the ``inverse" is performed on the
Liouville subspace spanned by $\mathcal{Q}$ where $\mathcal{L}$ is
regular (no null vectors).

Substituting the asymptotic behavior of the resolvent into Eqs.\
\eqref{Ffunction} and \eqref{Ffunctder}, keeping only the terms
divergent at $s=0$ in both equations, and performing the inverse
Laplace transform \footnote{It is important that
$F_{ii}(t;1)=\rho_{ii}(t)$ approach stationary state exponentially
fast so that they have no $1/t$ behavior for large times which
could combine in \eqref{noisegen} with the linear time divergence
of $\tfrac{\partial}{\partial z}F_{ii}(t;z)|_{z=1}$ to yield a
finite term.} we find the following large-$t$ asymptotics
\begin{equation}
\begin{split}
    \left.\begin{pmatrix}
    F_{00}(t;1)\\
    F_{11}(t;1)
    \end{pmatrix}\right|_{t\to\infty} &\to \mathcal{P}
    \begin{pmatrix}
    \rho_{00}^{\rm init}\\
    \rho_{11}^{\rm init}
    \end{pmatrix}=\begin{pmatrix}
    \rho_{00}^{\rm stat}\\
    \rho_{11}^{\rm stat}
    \end{pmatrix} \\
    \frac{\partial}{\partial z}\left.\begin{pmatrix}
    F_{00}(t;z)\\
    F_{11}(t;z)
    \end{pmatrix}\right|_{z=1,t\to\infty} &\to
    \begin{pmatrix}
    \rho_{00}^{\rm stat}\\
    \rho_{11}^{\rm stat}
    \end{pmatrix} \big(tI+C^{\rm init}\big)\\
    &\quad-
    \begin{pmatrix}
    \Sigma_{00}\\
    \Sigma_{11}
    \end{pmatrix},
\end{split}
\end{equation}
where we have defined an auxiliary quantity
$\Sigma=\mathcal{Q}\mathcal{L}^{-1}\mathcal{Q}\left(\begin{smallmatrix}
\Gamma_R e^{x/\lambda}\rho_{11}^{\rm stat}e^{x/\lambda}\\
0 \end{smallmatrix}\right)$ and $C^{\rm init}$ depends on initial
conditions. Using these in \eqref{noisegen} we arrive at the final
expression for the Fano factor $F=S(0)/2eI$
\begin{equation}\label{Fano}
    F = 1 - \frac{2e\Gamma_R}{I} \mathrm{Tr_{osc}}( e^{2x/\lambda}
    \Sigma_{11}).
\end{equation}
It is of crucial importance that this expression is independent of
the initial conditions [in the algebra leading to \eqref{Fano} the
linearly divergent terms in $t$ and the initial condition terms
cancel identically]. $\Sigma$ satisfies
\begin{equation}\label{GMRes}
    \mathcal{L}\Sigma=\begin{pmatrix}
    \Gamma_R e^{x/\lambda}\rho_{11}^{\rm stat}e^{x/\lambda}
    -\frac{I}{e} \rho_{00}^{\rm stat}\\
    -\frac{I}{e} \rho_{11}^{\rm stat}\end{pmatrix},\ \text{with }
    \mathrm{Tr_{sys}}\Sigma = 0.
\end{equation}
Equations \eqref{Fano}, \eqref{GMRes} together with the stationary
version of the GME
\begin{equation}\label{GMEstat}
   \mathcal{L}\begin{pmatrix}\rho_{00}^{\rm stat}\\\rho_{11}^{\rm stat}\end{pmatrix}=0,
   \ \text{with }\mathrm{Tr_{sys}}\rho^{\rm stat} = 1
\end{equation}
form the main theoretical result of this Letter and are the
starting point for the calculation of the noise properties of the
quantum shuttle.

In general, these equations have to be solved numerically.
However, there is an analytic solution to them in the limit of
small bare injection rates compared to damping, i.e.\
$\Gamma_{L,R}\ll\gamma\ll\om$. In this limit the oscillator gets
equilibrated between rare tunneling events and, consequently, the
state of the oscillator in a given charge state is close to its
corresponding canonical state, i.e. $\rho_{00}^{\rm
stat}=p_{00}\rho_{\rm osc}(0),\,\rho_{11}^{\rm
stat}=p_{11}\rho_{\rm osc}(eE),\, \rho_{\rm osc}(l)=e^{-\beta
(H_{\rm osc}-lx)}/\mathrm{Tr_{osc}}e^{-\beta (H_{\rm osc}-lx)}$
where only the probabilities $p_{00},p_{11}$ of the respective
occupations are to be determined from \eqref{GMEstat}. By tracing
Eq.~\eqref{GMEstat} with respect to the oscillator we find
$-\tilde{\Gamma}_L p_{00}+ \tilde{\Gamma}_R p_{11}=0$ (with
$p_{00}+p_{11}=1$) where $\tilde{\Gamma}_L=\Gamma_L
\mathrm{Tr_{osc}}(e^{-2x/\lambda}\rho_{\rm
osc}(0)),\,\tilde{\Gamma}_R=\Gamma_R
\mathrm{Tr_{osc}}(e^{2x/\lambda}\rho_{\rm osc}(eE))$ are the
renormalized tunneling rates. Proceeding similarly in the
evaluation of $\Sigma$ (this intuitive approach can be easily
justified with singular perturbation theory, see
e.g.~Ref.~\cite{vankampen}), we set $\Sigma_{00}^{\rm
stat}=s_{00}\rho_{\rm osc}(0),\,\Sigma_{11}^{\rm
stat}=s_{11}\rho_{\rm osc}(eE)$ and tracing Eq.~\eqref{GMRes} with
respect to the oscillator we arrive at $-\tilde{\Gamma}_L s_{00}+
\tilde{\Gamma}_R s_{11}= \frac{I}{e}\,p_{11}$ (with
$s_{00}+s_{11}=0$). Solving the equations for $p_{ii},\, s_{ii}$
we find $I = e\tfrac{\tilde{\Gamma}_L\tilde{\Gamma}_R}
{\tilde{\Gamma}_L+\tilde{\Gamma}_R}\ ,\ F =
\tfrac{\tilde{\Gamma}_L^2+\tilde{\Gamma}_R^2}
{(\tilde{\Gamma}_L+\tilde{\Gamma}_R)^2}$ which is the standard
result for the two-state sequential tunneling process
\cite{dav-prb-92,bla-phr-00}. The Fano factor only depends on the
ratio between the rates:
$\tfrac{\tilde{\Gamma}_R}{\tilde{\Gamma}_L}=\tfrac{\Gamma_R}{\Gamma_L}
\exp\tfrac{2eE}{\lambda m\om^2}$. This result provides us with the
analytic asymptotical expressions for the current and the Fano
factor for small hopping rates $\Gamma_{L,R}$ which we used to
check our numerical routine.

\begin{figure}
 \includegraphics[width=85mm]{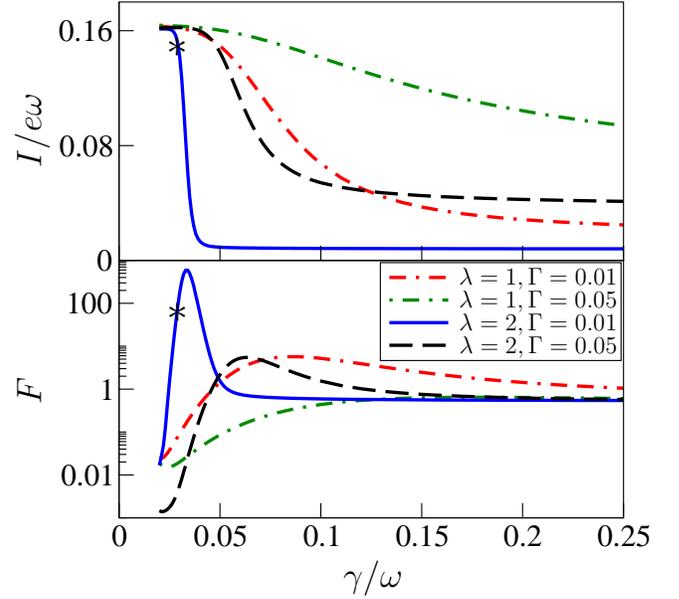}
  \caption{Current $I$ (upper panel) and Fano factor $F$ (lower panel; log scale)
  vs.\ damping $\gamma$ for different transfer rates $\Gamma$ and tunneling lengths
$\lambda$. The parameters are $\lambda=x_0,\Gamma=0.01\om$
(dot-dash-dashed line); $\lambda=x_0,\Gamma=0.05\om$ (dot-dashed
line); $\lambda=2x_0,\Gamma=0.01\om$ (full line);
$\lambda=2x_0,\Gamma=0.05\om$ (dashes) with
$x_0=\sqrt{\hbar/m\om}$. Other parameters are
$eE/m\om^2=0.5x_0,\,T=0$. The asterisk defines the parameters of
Wigner distributions in Fig.~\ref{wigner}.} \label{fano_fig}
\end{figure}

In the general case Eqs.~\eqref{GMEstat} and \eqref{GMRes} must be
solved by truncation of the oscillator Hilbert space by retaining
the $N$ lowest states of the oscillator, and solving numerically
the resulting $2N^2\times 2N^2$ linear systems. Since the required
$N$ for a satisfactory convergence could reach $N=100$ resulting
in big (non-sparse) linear systems we employed the Arnoldi
iteration and generalized minimum residual method (GMRes) for the
solution of Eq. \eqref{GMEstat} and \eqref{GMRes}, respectively
\cite{golub}. When properly implemented these methods provided a
fast solution with a modest memory requirement
\cite{fli-preprint-04}.

\begin{figure}
  \includegraphics[width=85mm]{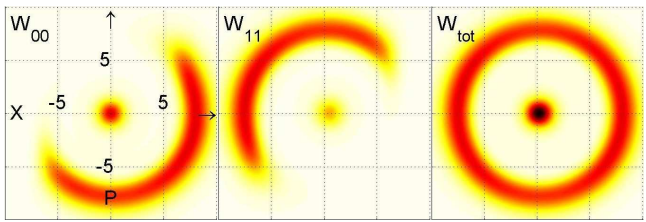}
  \caption{Phase space picture of the shuttle around the transition
  where the shuttling and tunneling regimes coexist.
  From left to right we show the Wigner distribution functions
  for the discharged ($W_{00}$), charged ($W_{11}$), and both ($W_{\rm tot}=W_{00}+W_{11}$)
  states of the oscillator in the phase space (horizontal axis --
  coordinate in units of $x_0=\sqrt{\hbar/m\om}$,  vertical axis -- momentum in $\hbar/x_0$).
  The values of the parameters are (corresponding to the asterisk in Fig.~\ref{fano_fig}):
  $\lambda=2x_0,eE/m\om^2=0.5x_0,\gamma=0.029\om,\Gamma=0.01\om,T=0$.}
  \label{wigner}
\end{figure}

In Fig.~\ref{fano_fig} we present the plots of the mean current
and the Fano factor as functions of the damping coefficient for
different parameters $\lambda$, and $\Gamma=\Gamma_L=\Gamma_R$.
The $I$ vs. $\gamma$ curve shows the tunneling to shuttling
crossover as damping is decreased, as explained in our previous
work \cite{nov-prl-03}. The crossover spans a narrower range of
$\gamma$'s in case of $\lambda=2x_0\ (x_0=\sqrt{\hbar/m\om})$
compared to the $\lambda=x_0$ case. Thus, already for
$\lambda=2x_0$ the shuttle behaves almost semiclassically, where a
relatively sharp transition between the two regimes is expected.
Around the transition the tunneling and shuttling regimes may
coexist, as shown analytically in \cite{fed-prl-04}. We see this
phenomenon explicitly in Fig.~\ref{wigner} where we plot the
Wigner distribution functions defined by
\begin{equation}
    W_{ii}(X,P) =
    \int_{-\infty}^{\infty}\frac{dy}{2\pi\hbar}\,\Bigl\langle
    X-\frac{y}{2}\,\Big|\rho_{ii}^{\rm stat}\Big|X+\frac{y}{2}\Bigr\rangle\,
    \exp\Bigl(i\frac{Py}{\hbar}\Bigr)
\end{equation}
for a specific set of parameters corresponding to the ``most
classical" curve around the transition in Fig.~\ref{fano_fig}
denoted by the asterisk. The Wigner plots show the
quasiprobability distributions in the phase space of the
oscillator resolved with respect to its charge state and prove the
coexistence of the tunneling regime (characterized by the spots
around the phase-space origin) and the shuttling regime with the
half-moon or ring-like shapes in $W_{00},W_{11}$ or $W_{\rm tot}$,
respectively \cite{nov-prl-03}. This semiclassical transition is
accompanied by the nearly singular behavior of the Fano factor
reaching the value $\approx 600$ at the peak. This is in agreement
with the recent classical study \cite{isa-epl-04} where the
singularity of the Fano factor at the transition was also
predicted.

More important, however, is the behavior in the shuttling regime.
We can see in Fig.~\ref{fano_fig} that the Fano factor is very
small in the shuttling regime. This is true even in the strongly
quantum case $\lambda=x_0$ where the transition peak
characteristic of the classical case is almost totally missing. As
found previously \cite{nov-prl-03} the classical transition is
strongly smeared by the quantum noise into a broad crossover which
is reflected by the absence of the peak in the Fano factor.
Nevertheless, the shuttling regime still persists and is again
characterized by a low noise.

To conclude, we have presented a generic method of the calculation
of the shot noise for quantum nanoelectromechanical systems and
applied it to a quantum shuttle system. We show that even in the
quantum case the shuttling regime is characterized by a highly
ordered charge transfer mechanism accompanied by the low current
noise compared to the tunneling regime. When approaching the
semiclassical limit, the Fano factor shows a giant enhancement at
the shuttling threshold, consistent with other classical studies
and with the phase space analysis of the stationary density
matrix.

Support of the grant 202/01/D099 of the Czech grant agency for
T.N. and of the Oticon Foundation for C.F. is gratefully
acknowledged. We thank T.~Brandes for attracting our attention to
the relevant literature and T.~Eirola for advice on numerics.

%\bibliography{Noise}

\end{document}